# Multiple EffNet/ResNet Architectures for Melanoma Classification

Chentian Ma[1, †], Li Li[2,†], Xuan Wen[3,†], and Jiaqi Xue[4,†]

[1]Bell Honors School, Nanjing University of Posts and Telecommunications, Nanjing, China
[2]School of Computer Science, Hubei University of Technology, Wuhan, China
[3]Computer Science and Technology, Chongqing University, Chongqing, China

[†]These authors contributed equally.

**Abstract:** Melanoma is the most malignant skin tumor and usually cancerates from normal moles, which is difficult to distinguish benign from malignant in the early stage. Therefore, many machine learning methods are trying to make auxiliary prediction. However, these methods attach more attention to the image data of suspected tumor, and focus on improving the accuracy of image classification, but ignore the significance of patient-level contextual information for disease diagnosis in actual clinical diagnosis. To make more use of patient information and improve the accuracy of diagnosis, we propose a new melanoma classification model based on EffNet and Resnet. Our model not only uses images within the same patient but also consider patient-level contextual information for better cancer prediction. The experimental results demonstrated that the proposed model achieved 0.981 ACC. Furthermore, we note that the overall ROC value of the model is 0.976 which is better than the previous state-of-the-art approaches.

## 1. Introduction

Skin cancer is the most widespread cancer diagnosed in the world. Moreover, the mortality rate of skin cancer is relatively high compared to other cancers. Specifically, melanoma is responsible for 75% of skin cancer deaths, despite being the least common skin cancer. Melanoma, usually called "malignant melanoma", is a malignant tumor that results from changes in melanocytes. Although melanoma is a typical skin disease, it can also occur in mucous membranes and internal organs. In recent years, the morbidity and mortality of melanoma have been increasing year by year, and it currently has no suitable treatment methods other than early surgical resection. Therefore, the early diagnosis of melanoma is particularly important. Hence it is necessary to get to know at the earliest whether the symptoms of the patient correspond to cancer or not.

There has been a lot of work published in the domain of skin cancer classification using deep learning and computer vision techniques. The works used a lot of different approaches including classification only, segmentation and detection, image processing using different types of filters etc. For example, Almansour et al. [1] developed an algorithm for the classification of melanoma by using k-means clustering, and Support Vector Machine (SVM). Abbas et al. [2], and Capdehourat et al. [3] separately used AdaBoost MC to classify skin lesions. Also, Isasi et al. [4] developed an algorithm for the diagnosis of melanoma. Past research on machine learning methods and their applications in pattern recognition systems were restricted to the transformation of a raw input into a representation of important handmade feature vectors which then, could be fed to a classifier for the classification and detection of patterns. However, in the past recent years with the exponential rise in  computation power and a large amount of data, deep learning techniques have become popular among researchers. For instance, Esteva et al. [5] made a breakthrough on skin cancer classification by a pre-trained

GoogleNet Inception v3 CNN model to classify 129,450 clinical skin cancer images including 3,374 dermatoscopic images. In addition, Yu et al. [6] developed a convolutional neural network with over 50 layers on ISBI 2016 challenge dataset for the classification of malignant melanoma. However, dermatologists evaluate every one of a patient's moles to identify outlier lesions that are most likely to be melanoma. Existing deep learning methods have not adequately considered this clinical frame of reference. Hence, we use ResNet and EffNet for better skin cancer classification.

As for the advantages of choosing ResNet model and EffNet model, ResNet can solve the problem that the effect of the deep network is decreased. The ResNet network can converge faster, and the residual structure can solve the problem of network degradation and gradient disappearance and explosion. EffNet is an open-source library that uses a new compound model scaling method and leverages recent progress in AutoML to improve neural network scaling techniques. By comparing and analyzing the experimental results, it can be found that the overall ROC value of the model is 0.976. This paper aims at realizing a deep learning approach for a multi-class classification by applying state-of-the-art pre-trained ResNet and EffNet on public skin cancer databases and FNN on patient-level contextual information which can yield higher diagnostic accuracy. Better detection of melanoma can positively impact millions of people.

The main contributions of this work can be summarized as follows:

1. The model makes a better prediction on skin cancer classification by comparing the results of ResNet and EffNet.

2. The model uses images within the same patient and patient-level contextual information for further providing assist on cancer prediction.

## 2. Model Formulation

The overall framework of the proposed model is shown in Fig. 1. The text and image data have been preprocessed and divided into the training set and the test set. The transfer learning of ResNet and EffNet is done before model training. For the image part, the image information is trained through ResNet50 and EffNet, and the model which has the best ROC has been obtained through comparison. Furthermore, FNN is used to process the text data. It should be noted that we also use the K-fold method to process non-overlapping data sets, and finally get the prediction model. The model can give the final prediction result of "Yes" or "No" for the input image and text information.

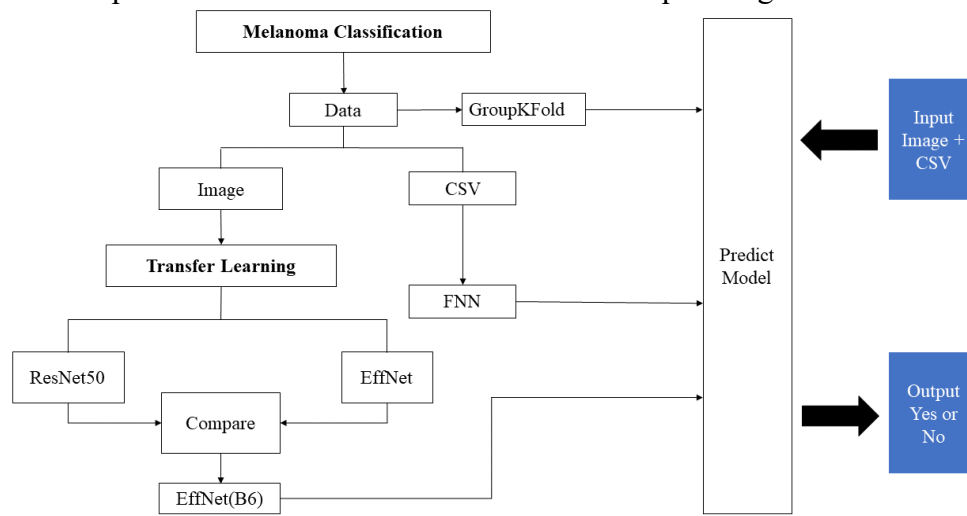

**Fig. 1** The overall framework of the proposed model.

## 2.1 Transfer Learning

Transfer learning in computer vision is the process of using a pre-trained model on a new dataset of images. This is useful when the data set is not particularly larger. ML models thrive when there is a lot of data available to train on, but in many scenarios, obtaining large datasets is quite difficult. The pre-trained model is trained on a greater dataset, and the weights are preserved. The trained layers of the model are used as the basis for our new model. We use the ResNet model and EffNet model to predict whether or not a lesion is a malignant or benign tumor.

## 2.2 The Structure of ResNet50

Residual Networks (ResNet) [7] are deep convolutional networks where the basic idea is to skip blocks of convolutional layers by using shortcut connections. The basic blocks named "bottleneck" blocks follow two simple design rules: (i) for the same output feature map size, the layers have the same number of filters; (ii) the number of filters is doubled if the feature map size is halved. ResNet with 50 layer network architecture that is used for classification is shown in Fig. 2.

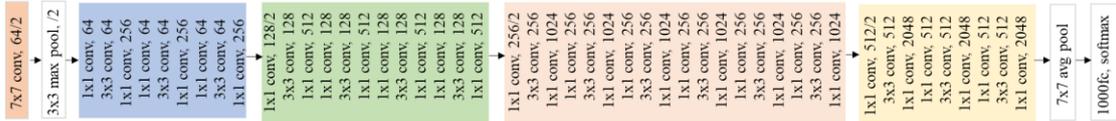

**Fig. 2** Step by Step model of ResNet50.

The portrayal strategy of arranged deep neural frameworks resembles that of a multilayer energy process. The image is used ad data to transmit layer by layer until the point when the moment that the characterizes result.

At the primary layer, ResNet50 utilizes 7x7 convolutions with stride 2 to down-sample the contribution by the request of 2, like the pooling layer. At that point, it's trailed by three characters obstructs before down-sampling again by 2. The own-sampling layer is additionally a convolution layer, yet without the personality association. It proceeds with like that for a few layers profound. The last layer is normal pooling which makes 1000 element maps for ImageNet information, and nourished into softmax layer straightforwardly; thus it's completely convolutional and at the end, we obtain the classification for the image that belongs to which class.

The activation function which is used in convolutional layers is rectified linear unit (Relu) which is a non-linear activation function.

$$f(y) = \max(y, 0) \tag{1}$$

At every readiness organize solitary center points are either dropped out of the net with probability p; therefore a lessened framework is left. Drawing nearer and dynamic edges to a dropped-out center point are moreover emptied. The dropout esteem is set to 0.5 that implies those hubs which have esteem under 0.5 are precluded from the preparation arrangement. At long last, the softmax work is utilized for the last expectation of the yield class.

$$f(x_i) = \frac{e^{x_i}}{\sum_{j=0}^{K} e^{x_j}}, i = 0, 1, 2, \mathrm{K}\ K \tag{2}$$

The softmax function finds out the probabilities of every target class over all possible target classes. Later the processed probabilities are helpful for choosing the target class for the given sources of info. It ranges from 0 to 1 and the total of the extensive number of probabilities are proportional to one.

## 2.3 The Structure of EfficientNet

Convolutional neural networks are usually developed under a fixed resource budget. If more resources are available, they will be expanded to obtain better accuracy. For example, the network depth, network width, and input image resolution can be increased. However, it is very difficult to manually adjust the depth, width, and resolution to enlarge or reduce. When the amount of calculation is limited, it is difficult to determine which one to enlarge or reduce. In other words, the combined space is too large. Manpower cannot be exhausted. Based on the above background, this paper proposes a new model scaling method, which uses a simple and efficient composite coefficient to enlarge the network from the three dimensions of depth, width, and resolution, instead of arbitrarily scaling the dimensions of the network like traditional methods. Based on neural structure search technology, an optimal set of parameters (composite coefficients) can be obtained. As can be seen from the figure below, EfficientNet [8] is not only much faster than other networks, but also has higher accuracy.

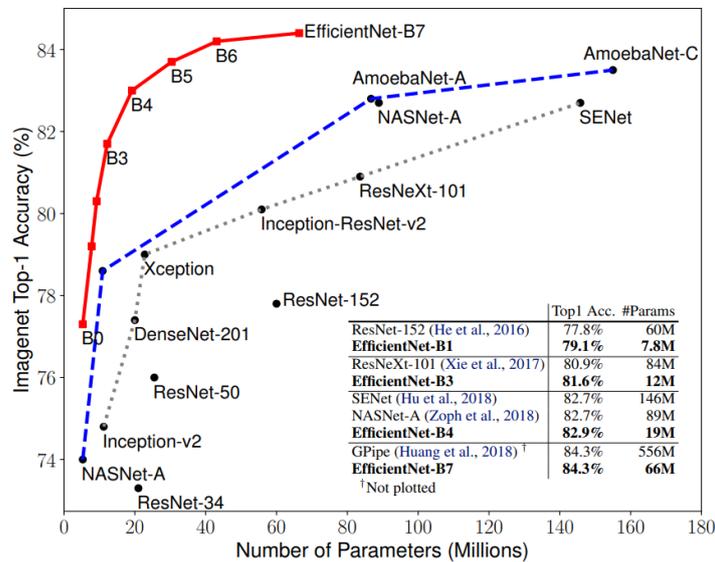

**Fig. 3** Multiple EffNet model

By enlarging the basic model of EfficientNet, a series of EfficientNet models are obtained. This series of models defeated all previous convolutional neural network models in terms of efficiency and accuracy. In particular, EfficientNet-B7 obtained the top-1 accuracy rate of 84.4% and the top-5 accuracy rate of 97.1% on the ImageNet data set. And compared with other models with the highest accuracy at the time, the size was reduced by 8.4 times and the efficiency was increased by 6.1 times. And through transfer learning, EfficientNet reached the most advanced level at the time on multiple well-known data sets.

In this case, we choose the basic network model EfficientNet-B0 in the EfficientNet series. When the model is trained on the ImageNet dataset, it contains a total of 5,330,564 parameters, of which 5,288,548 parameters are required for gradient descent training. The parameters that do not need to be trained are 42016 in the mean and variance in the Batch Normalization layer. The core structure of the network is the mobile inverted bottleneck convolution (MBConv) module, which also introduces the attention thought of the compression and excitation network (Squeeze-and-Excitation Network，SENet). The highest accuracy rate at the time was achieved on the ImageNet data set.

The mobile flip bottleneck convolution is also obtained through the neural network architecture search. The module structure is similar to the depthwise separable convolution. The mobile flip bottleneck convolution first performs a 1*1 point-by-point convolution on the input and based on the expansion ratio (expand ratio). Change the output channel dimension (for example, when the expansion ratio is 3, the channel dimension will be increased by 3 times. But if the expansion ratio is 1, then directly omit the 1*1 point-by-point convolution and subsequent batch normalization and activation functions). Then proceed to the depthwise convolution of k*k. If compression and excitation operations are to be introduced, the operations will be performed after deep convolution. Then restore the original channel dimension at the end of 1*1 point-by-point convolution. Finally, the drop connects and skip connection of the input are performed. This approach is derived from the paper "Deep networks with stochastic depth", which gives the model a random depth and shortens the model training needs Time, improve the model performance (note that in EfficientNet, only when the same mobile flip bottleneck convolution recurs, will the connection deactivation and the input skip connection be performed, and the deep convolution step Long becomes 1), connection inactivation is an operation similar to random dropout, and an identity jump is added at the beginning and end of the module. Note that after each convolution operation in this module, batch normalization is performed, and the activation function uses the Swish activation function.

The network is divided into 9 stages in total. The first stage is an ordinary convolutional layer with a convolution kernel size of 3x3 and a step size of 2 (including BN and activation function Swish). Stage2~Stage8 are all repeatedly stacking the MBConv structure (last The Layers in a column indicates how many times the Stage repeats the MBConv structure), and Stage9 consists of a common 1x1 convolutional layer (including BN and activation function Swish), an average pooling layer and a fully connected layer. Each MBConv in the table will be followed by a number 1 or 6, where 1 or 6 is the magnification factor n, that is, the first 1x1 convolutional layer in MBConv will expand the channels of the input feature matrix by n times, where k3x3 or k5x5 Represents the size of the convolution kernel used by Depthwise Conv in MBConv. Channels represent the Channels that output the feature matrix after passing through the Stage.

### 2.4 Combination of FNN and CNN

We use the ResNet50 model mentioned above to process the images in the dataset and use the FNN model to process csv data in the dataset.

And connecting the results processed by the ResNet50 model and the FNN model for classification shown in Fig. 4.

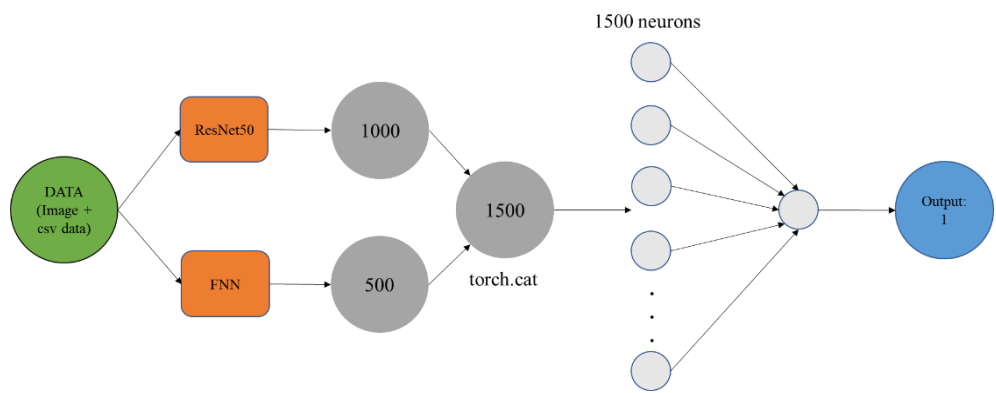

**Fig. 4** Combination of FNN and ResNet

In addition, we also use the EffNet model mentioned above to process the images in the dataset and use the FNN model to process csv data in the dataset.

And also connecting the results processed by the EffNet model and the FNN model shown in Fig. 5 for classification.

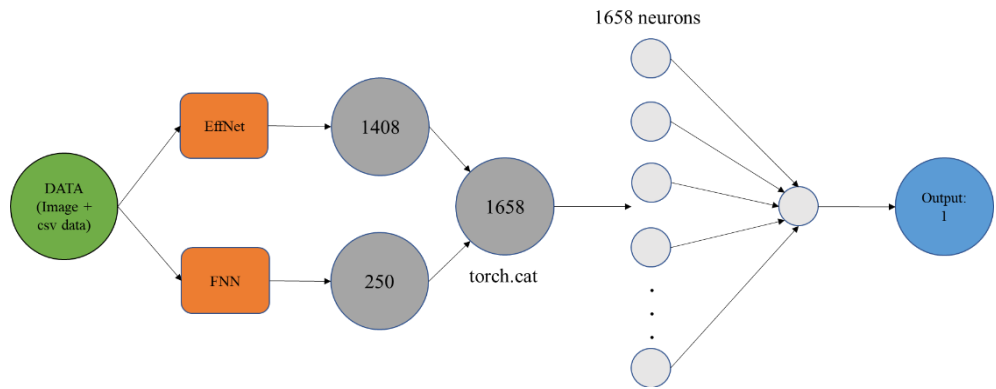

**Fig. 5** Combination of FNN and EffNet

In the ResNet model and EffNet model, we use the BatchNorm1d function to perform batch standardization operations on small batches of input. And fully connected layer plus dropout layer to prevent the model from overfitting and improve model generalization ability.

**2.5 GroupKFold**

The same group will not appear in two different folds (the number of distinct groups has to be at least equal to the number of folds).

We're using patient_id for our grouping column: there are multiple patients with multiple images taken, so we need to be careful with that. The realization idea of GroupKFold is shown in Fig. 6.

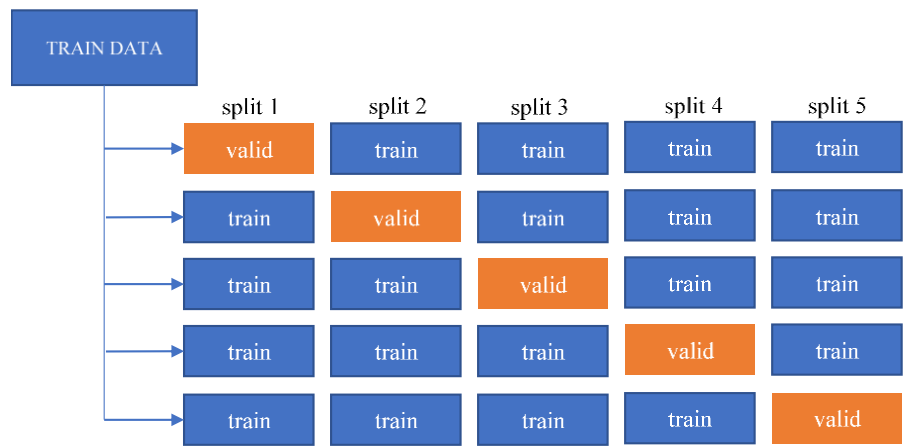

**Fig. 6** The structure of GroupKFold

## 3. Experiments

All experiments are conducted on a computer with 2.60GHz Intel i7-9750H processor and 16GB RAM under Windows 10 operating system. The program codes of data preprocessing and graphs modeling are written by Python 3.7.

**3.1 Datasets**

The datasets are taken from the Society for Imaging Informatics in Medicine (SIIM), as the leading healthcare organization for informatics in medical imaging, the SIIM 's mission is to advance medical imaging informatics through education, research, and innovation in a multi-disciplinary community. SIIM is joined by the International Skin Imaging Collaboration (ISIC), an international effort to improve melanoma diagnosis. The ISIC Archive contains the largest publicly available collection of quality-controlled dermoscopic images of skin lesions.

The details of the datasets utilized in the experiments are illustrated as follows.

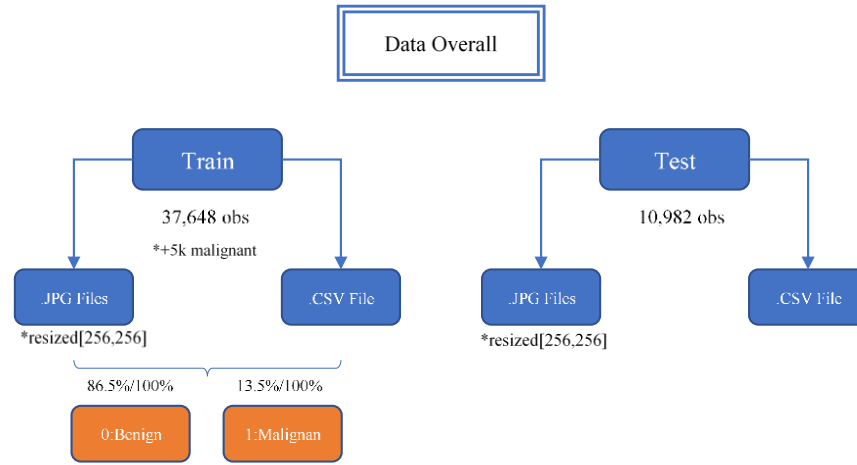

**Fig. 7** Data Distribution

The images are provided in DICOM format. This can be accessed using commonly available libraries like pydicom, and contains both image and metadata. It is a commonly used medical imaging data format.

Images are also provided in JPEG and TFRecord format (in the jpeg and tfrecords directories, respectively). Images in TFRecord format have been resized to a uniform 1024x1024. Metadata is also provided outside of the DICOM format, in CSV files.

**Tab.1** Attributes in CSV files

| image_name | unique identifier, points to filename of related DICOM image |
|---|---|
| patient_id | unique patient identifier |
| sex | the sex of the patient (when unknown, will be blank) |
| age_approx | approximate patient age at the time of imaging |
| anatom_site | location of the imaged site |
| diagnosis | detailed diagnosis information (train only) |
| benign_malignant | indicator of malignancy of the imaged lesion |
| target | a binarized version of the target variable |

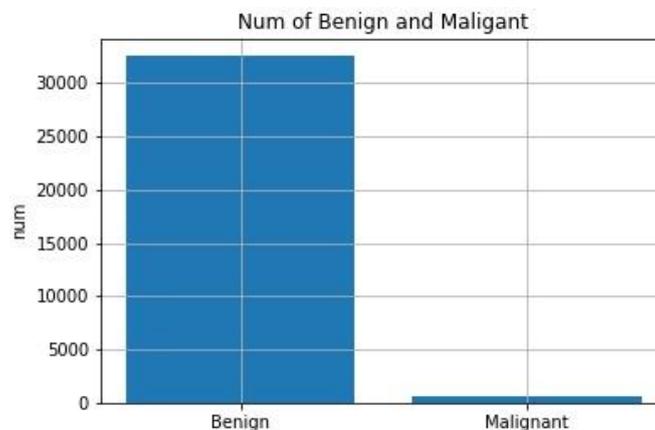

**Fig. 8** The proportions of the two types of image data sets

This is a very important topic in this classification problem, as the 2 classes we are dealing with are highly imbalanced, with 98% of the data being benign and only 2% of the data being malignant. This is also the kind of problem where you don't want to have False Negatives. It's worse to tell a patient they don't have cancer when they actually do, than to tell they do have it and they actually don't. So, having balanced classes is crucial. This experiment takes an approach to solve this problem: oversampling or undersampling. The oversampling is, of the minority class, increasing the number of images through augmentations.

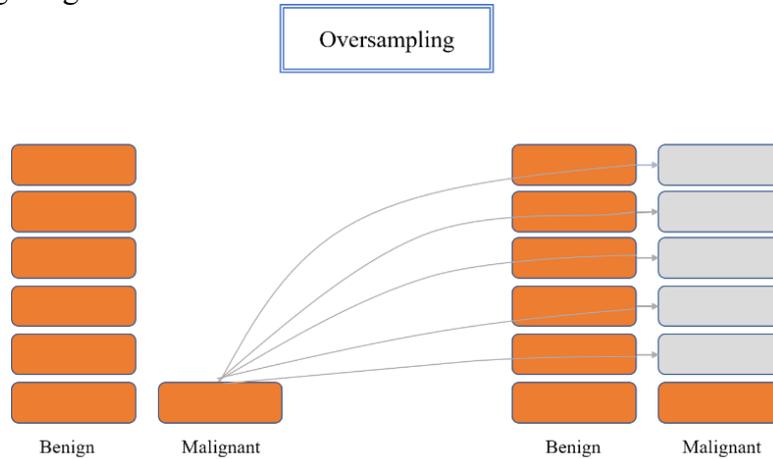

**Fig. 9** Overview of Oversampling

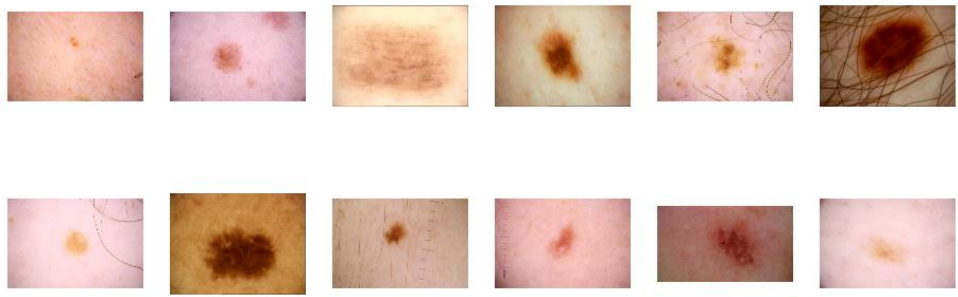

**Fig. 10** The original image of the image dataset

In the experiment, we want to eliminate the influence of illumination, size, contrast and saturation on recognition. We use a library that goes hand in hand with PyTorch and it's easily used to augment data.

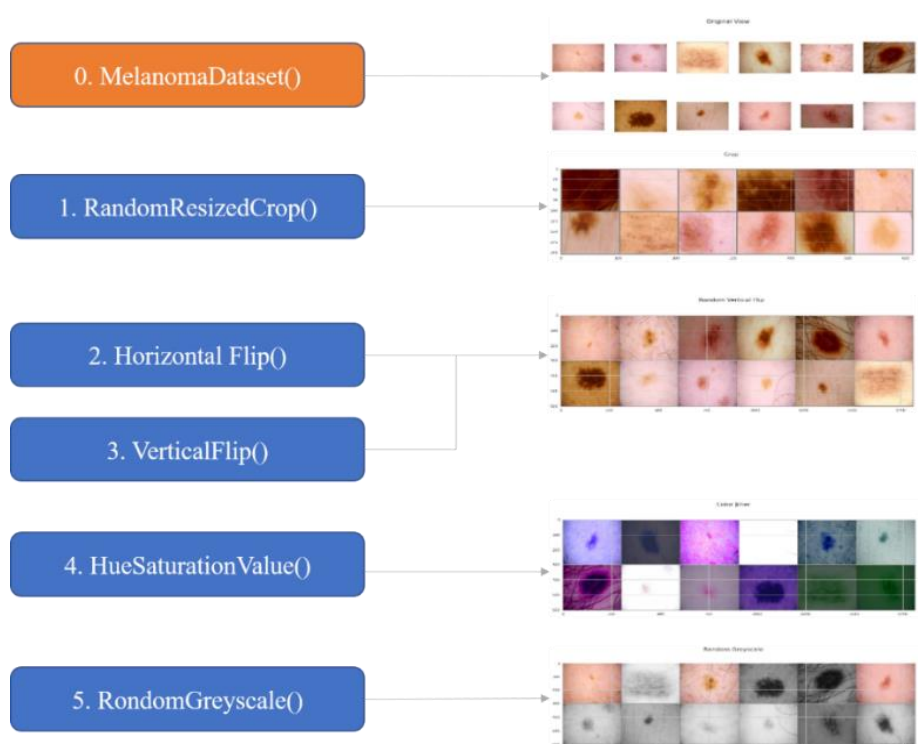

**Fig. 11** The process of Augmentations

1. First, crop the pictures into the same size.
2. Vertically flip and horizontal flip the given PIL Image randomly with a given probability.
3. Then randomly change the brightness, contrast and saturation of an image.
4. Further, we randomly convert the image to grayscale with a probability of p (default 0.1).

As you may have noticed, all images are for light skin colors, so no preprocessing in this area is required. However, hair removal might be a good augmentation that will help the model perform better. We have realized the hair removal method with the help of CV2. The following is the effect achieved.

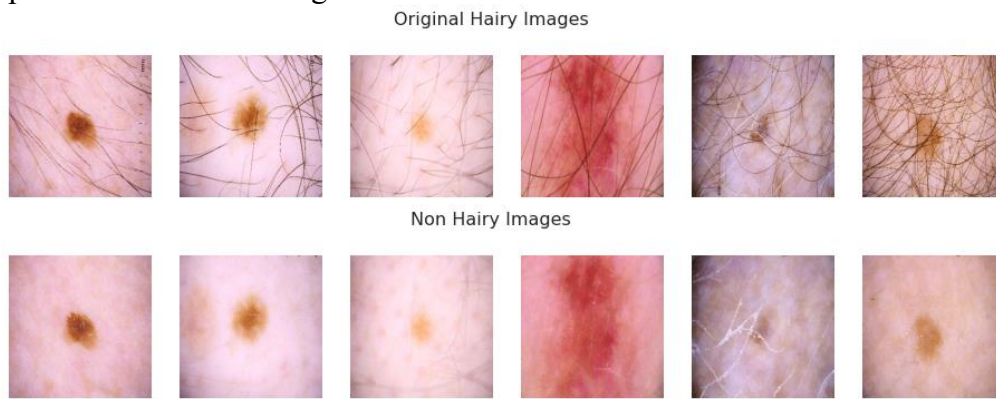

**Fig. 12** Hair cut technique implement in the image datasets

This is the statistical chart of the distribution of malignant tumors on the body.

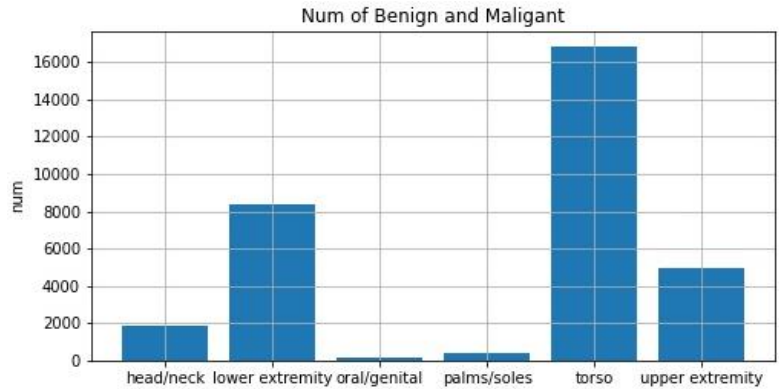

**Fig. 13** Distribution statistics of skin tumors

It seems like people have a lot of issues with the torso, and after that the extremities of the body (upper/lower). We have about 100-200 cases of cancer in the mouth or genitalia (the areas with the lowest rate of cancerous growth) and the palms and soles also are safe (probably because they are not exposed to any external sources in the day-to-day life of the person).

One could indeed fathom that the torso is frequently exposed either during the occasional workout or the occasional swim, or in some cases, the occasional extreme adventure. It also could be that the torso was exposed to UV light (which is the cause of melanoma) in highly populated regions (where pollution allows the sun's UV rays to come in). In addition to the picture, other attributes of the patient are also very important, such as age, location, and gender; so, we should transform all categorical features numerically. After doing this, the data preprocessing is all completed.

### 3.2 Evaluation Metrics

**Confusion matrix**

A confusion matrix contains information about actual and predicted classification done by a classification system. The performance of such systems is commonly evaluated using the data in the matrix. The confusion matrix in the melanoma classification problem is shown in Fig. 14.

|  |  | Predicted class | |
|---|---|---|---|
|  |  | Benign | Malignant |
| Actual class | Benign | $a$ | $b$ |
|  | Malignant | $c$ | $d$ |

**Fig. 14** Confusion matrix in a melanoma classification problem

- $a$ is the number of people who did not have cancer and were also predicted to not have cancer

- $b$ is the number of people who did not have cancer but the prediction says they do

- $c$ is the number of people who have cancer but the predictions say they don't

- $d$ is the number of people who have cancer and were also predicted to have cancer.

A few standard terms are defined in the matrix: accuracy, true positive rate, false positive rate, true negative rate, false negative rate and precision.
True positive rate is the proportion of positive cases that are properly identified and can

be calculated using the equation:

$$\text{True positive rate} = \frac{d}{c+d} \qquad (3)$$

The false positive rate is the proportion of negative cases that were incorrectly classified as positive, and calculated with the equation:

$$\text{False positive rate} = \frac{b}{a+b} \qquad (4)$$

The true negative rate was defined as the proportion of negatives cases which are classified correctly, and is calculated using the equation.

$$\text{True negative rate} = \frac{a}{a+b} \qquad (5)$$

The dales negative rate is the proportion of positive cases that were incorrectly classified as negative, and are calculated using the equation:

$$\text{False negative rate} = \frac{c}{c+d} \qquad (6)$$

Fig. 14 shows that there are four possible outputs which represent the elements of a 2x2 confusion matrix or a contingency table. If the sample is benign and it is classified as benign, it is counted as a true positive (TP); if it is classified as malignant, it is considered as a false negative (FN) or Type II error. If the sample is malignant and it is classified as benign, it is considered as true negative (TN); if it is classified as benign, it is counted as false positive (FP), false alarm or Type I error. As we will present in the next sections, the confusion matrix is used to compare the classification performance of different models.

**The area under the ROC curve (AUC)**

AUC (area under the curve) is the area under the ROC curve. The calculation formula of AUC area is shown in Eq. (7).

$$AUC = 1 - \frac{1}{m^+ m^-} \sum_{x^+ \in D^+} \sum_{x^- \in D^-} W(f(x^+) < f(x^-)) + \frac{1}{2} W\left(f(x^+) = f(x^-)\right), \qquad (7)$$

where $m^+$ is the number of positive samples, and $m^-$ is the number of negative samples. $D^+$ is the set of all positive examples, $x^+$ is one of the positive examples, $D^-$ is the set of all negative examples, $x^-$ is one of the negative examples, $f(x)$ is the result of model prediction for the sample $x$, and the value is between $[0,1]$, $W(x)$ takes the value 1 only when $x$ is true, otherwise it takes the value 0.

We need to use the two variables defined above to build ROC diagram: *False positive rate* (FPR), *True positive rate* (TPR). If you take FPR as the x-axis and TPR as the y-axis, you can get a coordinate system shown in Fig. 15.

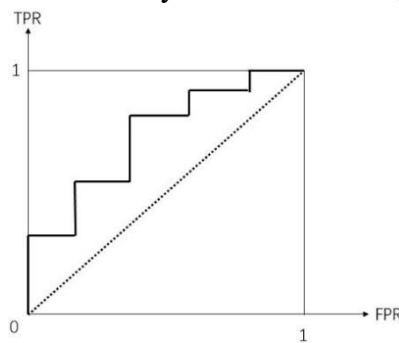

**Fig. 15** ROC diagram example

The dashed line in Fig. 15 is our hypothetical prediction result. The ROC curve drawn is generally above the dotted line, because the result is generally better than the average. The bold curve in the figure is the ROC curve, and the area under the ROC curve is the AUC.

We can use the following explanation to understand the quality of the prediction results. Point (0,1), that is, FPR=0, TPR=1. According to the formula mentioned above, if FPR=0, then FP=0, which means there are no false positives. If TPR=1, then FN=0, which means there is no false counterexample. This is the perfect situation; all predictions are correct. The benign is predicted to be benign, and the malignant is predicted to be malignant, and the classification is 100% correct. This reflects the significance of FPR and TPR, so we hope that the smaller the FPR, the better, and the larger the TPR, the better. Point (1,0), that is, FPR=1, TPR=0. This point forms a contrast with the point (0,1), just the opposite. So, this is the worst-case scenario, because all predictions are wrong. Point (0,0), that is, FPR=0, TPR=0. Then FP and TP=0. So, the significance of this point is that all samples are predicted to be malignant. In other words, no matter what sample is given to me, it is predicted to be malignant. Point (1,1), that is, FPR=1, TPR=1. This point is in contrast with the point (0,0). The meaning of this point is to predict all samples as benign.

After understanding these four points, we can know that the closer a point is to the upper left corner, the better the prediction effect of the model. The larger the value of AUC, the higher the prediction accuracy of the classification model. If it can reach the upper left corner, which is the point (0,1), that is a perfect result.

## 3.3 Experimental Results and Analysis

### 3.3.1 Experiment 1: Comparison between ResNet and EffNetB6

We apply ResNet and EffNetB6(fine-tune version) to the training set and validation set respectively, and compare the convergence speed, accuracy and recall rate of the two CNN models.

Fig.16 and Fig.17 show the training curve of ResNet and EffNetB6. In the training set, by comparing the curve of the loss function, it can be observed that EffNetB6 starts faster and converges faster; in the same number of iterations, the accuracy of EffNetB6 is higher, and it can reach 0.72 in the first iteration; in the recall rate of benign samples, the performance of EffNetB6 is slightly better than ResNet, but in the recall rate of malignant samples, there is a significant difference between EffNetB6 and ResNet, the performance of EffNetB6 is obviously better.

But at the same time, we also notice that although EffNetB6 has the characteristics of faster convergence, higher accuracy and recall in the training set, EffNetB6 is far less stable than ResNet in the validation set, which will affect the reliability of EffNetB6 in practical application to a certain extent.

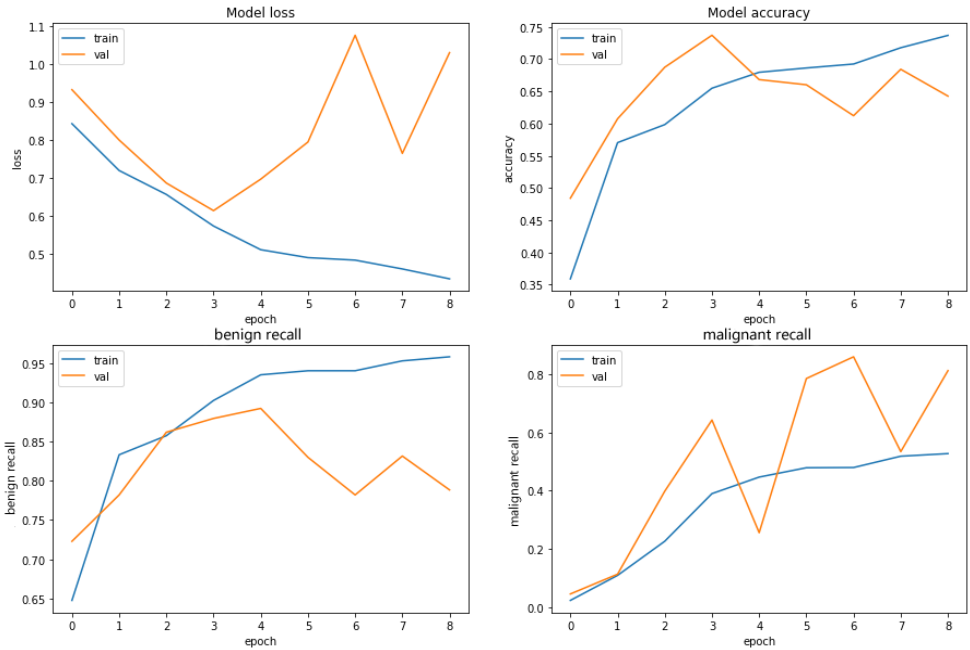

**Fig. 16** Loss function curve, accuracy curve and recall rate curve of ResNet (fine-tune version)

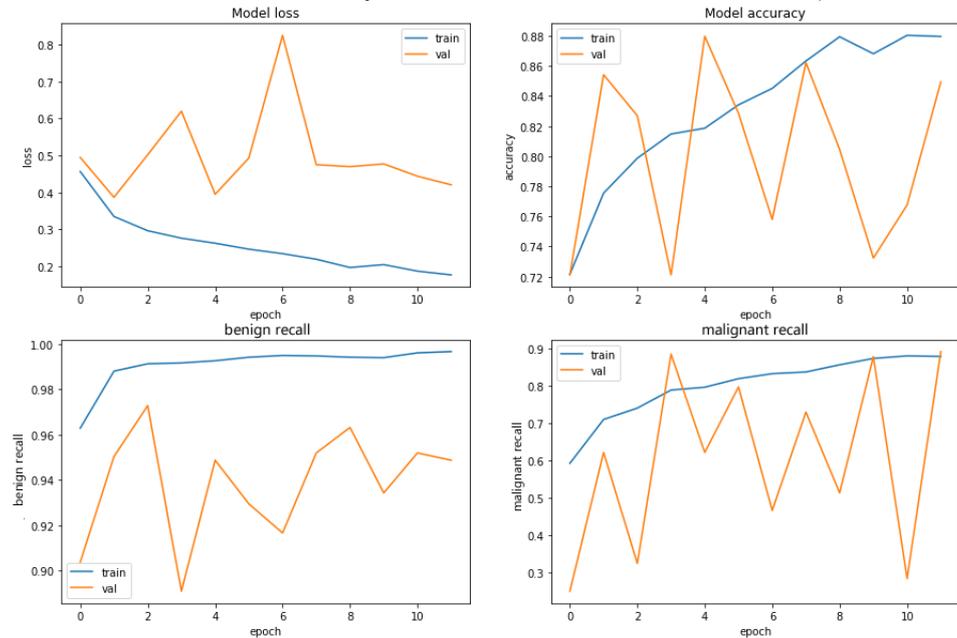

**Fig. 17** Loss function curve, accuracy curve and recall rate curve of EffNetB6 (fine-tune version)

For this problem, we make the following assumptions: (1) this is due to the overfitting problem in the validation set; (2) because we choose to freeze the model parameters (fine tune) and generalize them directly to our application. Therefore, because we find that the performance of the model is not ideal, based on assumptions, we realize that this is because the weight of the image network is not suitable for this kind of image classification. Because of the better performance of EffNetB6 in the above experiments, we choose EffNetB6 as the follow-up experimental object. We set the EffNetB6 layer as trainable to train the whole model for better performance.

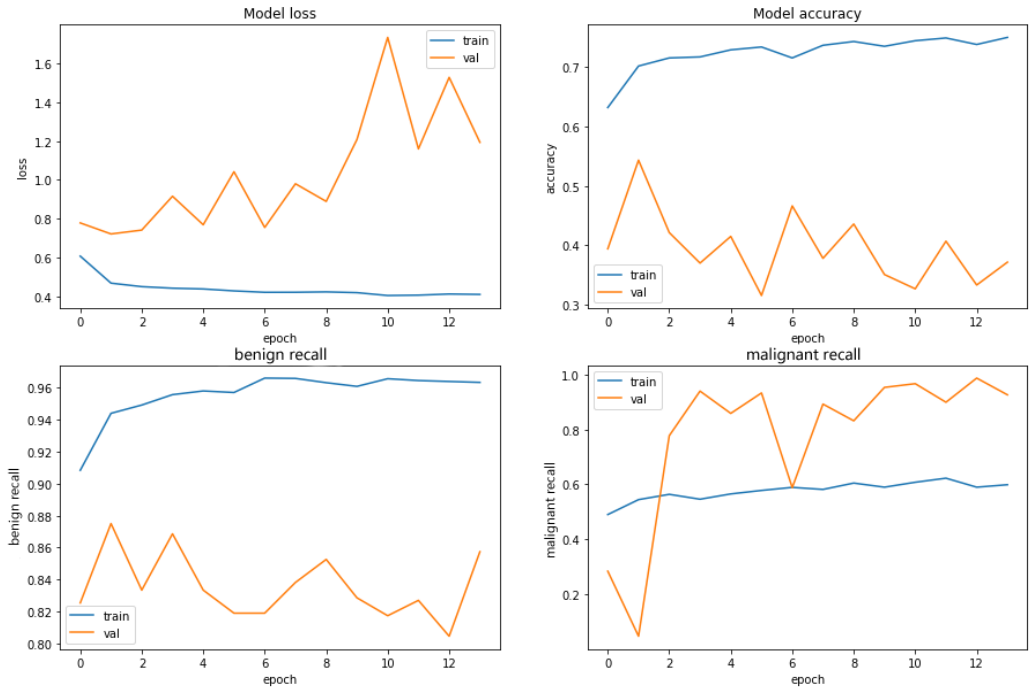

**Fig. 18** Loss function curve, accuracy curve and recall rate curve of EffNetB6
(open frozen layer and add dropout method to prevent overfitting)

From the experiment without opening the frozen layer, we observed that EffNetB6 is better and faster than ResNet. After opening the frozen layer in EffNetB6, the training effect is obviously better, and the performance on the verification set is more stable, which confirms the previous assumptions. Even in the recall rate of malignant samples, EffNetB6 performs better in the validation set. Although the convergence rate is much slower after the frozen layer is opened, the possible reason is that the training features are more accurate after the frozen layer is opened.

<!-- Confusion Matrix: OOF Data -->

|  | benign | malignant |
|---|---|---|
| **benign** | True Neg 32,344 85.9% | False Pos 198 0.5% |
| **malignant** | False Neg 877 2.3% | True Pos 4,229 11.2% |

**Fig. 19** confusion matrix of final model
(all layers of EffNetB6 are opened and dropout is added to prevent overfitting)

From the confusion matrix (shown in Fig. 19), the proportion of false positive samples was 0.5%, and the proportion of false negative samples was 2.3%. The TPR was 0.830, FPR was 0.00579, and ROC was 0.976. The model performed well and had good accuracy and reliability.

### 3.3.2 Experiment 2: Comparison between CNN only and CNN+FNN

Traditional skin cancer classification only uses convolutional neural network analysis of photos. However, other patient information is also particularly important for cancer diagnoses, such as age, gender, and tumor location. Therefore, we used a combination of CNN and FNN, and we compared the performance of using CNN alone to analyze photos and using a combination of CNN and FNN to analyze photos and discrete attributes, respectively.

It can be concluded from the performance comparison graph that the recall rate of cancer judgments has been improved after adding an FNN for analyzing discrete attributes. It is reflected in the training set and the validation set.

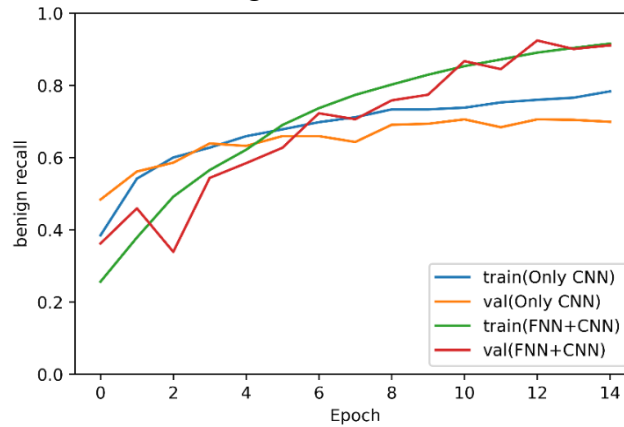

**Fig. 20** Comparison of the training process of CNN and FNN+CNN

## 4. Conclusion

This paper proposed a melanoma classification model based on EffNet and ResNet to achieve better use of patient information and improve the diagnostic accuracy. Our model not only uses images of suspected tumors for training, but also takes patient-level context information into account. Meanwhile, this paper compares the advantages of EffNet and ResNet convolutional neural networks in terms of the ROC value of the entire train data. The experimental results have shown that the model is superior to the existing methods in accuracy.

In the future, this work will focus on optimizing datasets and improving models. For example, more clinical patient information and the priority of these characteristics can be considered in the model. The past medical history and living habits of patients should be taken into account in the model. Furthermore, ResNet and EffNet can be combined for training in a large ensemble model to obtain a better performance in comparison with the current work.